\documentclass[aps,prb,showpacs,twocolumn,preprintnumbers,amsmath,amssymb]{revtex4}
\usepackage{dcolumn}% Align table columns on decimal point
\usepackage{bm}% bold math
\usepackage{amsmath}
\usepackage{feynmp}

\usepackage{graphicx}% Include figure files

\begin{document}

\title{Robust  topological insulator  conduction  under  strong boundary disorder}

\author{Quansheng Wu}
\thanks{Quansheng Wu and Liang Du contributed equally to this work.}
\affiliation{Institute of Physics, Chinese Academy of Sciences, Beijing 100190, China}

\author{Liang Du}
\affiliation{Institute of Physics, Chinese Academy of Sciences, Beijing 100190, China}

\author{Vincent E. Sacksteder IV}
\email{vincent@sacksteder.com}
\affiliation{Institute of Physics, Chinese Academy of Sciences, Beijing 100190, China}

 \pacs{72.25.-b,73.25.+i,72.15.Rn, 71.70.Ej }

\date{\today}% It is always \today, today,

\begin{abstract}
Topological insulators are characterized by  specially protected conduction  on their outer boundaries.  We show that the protected edge conduction exhibited by 2-D topological insulators (and also Chern insulators) is independent of non-magnetic boundary disorder.  In particular, the edge states residing inside the bulk gap remain conducting even when edge state inhomogeneities destroy the characteristic linear Dirac relation between energy and momentum.  The main effects of boundary disorder on the in-gap states are to decrease the Fermi velocity,  increase the density of states, pull the states into the disordered region if spin is conserved, and at very large disorder shift the states to the boundary between the disordered edge and the clean bulk.  These effects, which  may be useful for device engineering, are controlled by  a resonance between the disordered edge and the bulk bands.  The resonance's energy is set by the  bulk band width; protection of the in-gap edge states' plane-wave character is controlled by the bulk band width, not the bulk band gap.  
\end{abstract}

\maketitle

\section{Introduction}

Recently a new kind of material has been predicted and measured: topological insulators, which do not permit current to flow through their interior but do allow  metallic conduction along their boundaries.  \cite{Kane05, Zhang09, Hasan10, Li12, Culcer12}  The conducting boundary states are protected topologically, meaning that they are safeguarded by the bulk's insulating property from any finite-size perturbations as long as a mobility gap is maintained between the bulk bands. \cite{Zhang12,Xu12}  The boundary states  obey a linear Dirac dispersion like that seen in graphene, but unlike graphene their spin degeneracy is strongly broken,   the angles of the spin and the momentum are locked to each other, and $T$ symmetry prohibits any gap in the Dirac dispersion.  Topological insulators (TIs) have won tremendous  theoretical and experimental interest because these special properties have  potential for spintronics and quantum computation.  However TIs  continue to face critical experimental challenges.  Chief among these are elimination of residual bulk carriers from dopants introduced by the TI growth process and by exposure to air,   control of the Fermi level which responds strongly to doping, and fabrication of clean and stable surfaces.

This present article focuses on how non-magnetic disorder located \textit{only on the TI's boundary} affects topological protection. Several experiments on 3-D TIs have shown that TI surface states are not destroyed even under exposure to air. \cite{Analytis10, Brahlek11, Aguilar12, Tereshchenko11}  However three recent angularly resolved photoemission spectroscopy (ARPES) experiments measured the surface band while progressively doping the TI surface with impurities and found that the surface signal becomes progressively fainter and more blurred as impurities are added. \cite{Hsieh09,Noh11,Liu12}    Moreover several recent experiments have modified the TI surface by introducing an Al capping layer, or by gating the TI. \cite{Lang11, Chen10}  Another experiment reported that exposure to air causes growth of an oxide layer on top of the TI. \cite{Kong11}  Therefore a better understanding is needed of how topological protection of the states on a TI's boundary is affected by disorder located on the same boundary.  

Here it is important to distinguish carefully between experimental observables.  Many experiments have characterized  3-D TIs using ARPES, which   measures the surface states' spectral density -  their momentum-resolved  density of states.  At zero disorder  these states are plane waves with definite momenta and linear dispersion $E = \pm v_F |\vec{p}|$, and ARPES observes  a direct image of the linear Dirac cone.   We will show that measurements of the momentum-space structure of states on a TI's boundary,  such as the ARPES and Schlubnikov-de Haas techniques, are quite sensitive to disorder.  Since disordered electronic states are not plane waves, the observed energy dispersion is blurred at weak disorder and may be entirely absent at strong disorder.     However we will find that the  boundary states can remain topologically protected even when disorder removes their  plane wave character, and may manifest  themselves in other observables such as the conductance and magnetoconductance.  

The current literature offers  two lenses  for understanding  topological protection.  The first lens is focused on  bulk properties which protect the boundary states.  In TIs  the bulk  valence and conduction bands are  split both by a strong spin-orbit interaction and by a gap  whose width is controlled by a mass parameter $m$.   If the mass  and spin-orbit terms have opposite signs,  the conflict between the two signs entangles the valence band with the conduction band, and the states of each band acquire a special texture centered on the $\vec{k} = 0$ $\Gamma$ point. \cite{Bernevig06a}  This situation, called band inversion, strongly affects the TI's electronic structure both at its boundaries with the vacuum and also at its boundaries with any material having a positive (non-inverted) mass.   On these boundaries the bulk band inversion creates  a band of boundary states which connects the valence and conduction bands with each other and therefore bridges the bulk gap.  If the Fermi level lies in the bulk gap, then any states in the bulk decay exponentially and   current flows only in the in-gap  boundary states caused by the band inversion.   

The in-gap  states are safeguarded - topologically protected - by the bulk band gap.   Destruction of the boundary band can occur only when boundary states tunnel from one side of the sample to the other, i.e. when delocalized bulk states occur at the same energy as the boundary states.  Such bulk delocalization could be caused by the sample's geometry, \cite{Buttner11, Kvon11}  by interactions, \cite{Pesin10} or by disorder-assisted percolation through the bulk.  \cite{Chen12, Xu12}  Whatever its source, bulk delocalization  is not possible unless the bulk gap is destroyed throughout the sample, which is not possible  when only boundary disorder is present.   In particular,  boundary disorder leaves the bulk band gap  unchanged.  Therefore  we should expect in-gap boundary states to be robust against boundary disorder.  If the boundary disorder is strong enough to locally change the band structure in the disordered region, then the TI boundary states will simply detour around the disorder and ride the boundary between the disordered region and the clean TI bulk. \cite{Ringel12, Schubert12}

The second lens for understanding topological protection focuses  on  individual elastic scattering events which interrupt intervals of plane wave motion.    The boundary disorder is assumed to be weak, bulk states are neglected, and much importance is given to the plane wave boundary states which exist at zero disorder.  Especially important is the fact that their spin and momentum are locked together.    In a 2-D TI at fixed Fermi energy $E_F$ there are only two edge states which have opposite spin quantum number and opposite momenta $|\vec{p}| = \pm E_F / v_F$, where $v_F$ is the Fermi velocity.  Non-magnetic disorder is unable to flip the spin and therefore scattering between the two states is impossible; the edge conductance is always quantized at $2 G_0$, where $G_0$ is the conductance quantum.   This is a beautifully simple microscopic explanation  of how a 2-D TI's topological structure plays out in individual scattering events and guards edge conduction from weak disorder.     3-D TIs are a bit more subtle, since the zero disorder plane waves allow scattering at any angle except 180 degrees.  However here also  spin-momentum locking implies that electrons pick up a $\pi$ Berry phase whenever their scatterings form a closed loop, and this Berry phase  reverses the sign of  weak disorder's effects on the conductance; the conductance increases rather than decreases.  \footnote{This effect is typical in systems with strong spin-orbit coupling. \cite{Hikami80} } Therefore we see again in 3-D TIs that the conductance is guarded against weak boundary disorder.  This result has been extended by non-perturbative numerical studies of   conduction in a single spin-momentum locked Dirac cone  band on a 2-D surface without an adjoining bulk, which indicate that the conductance increases without limit, proportional to $\ln L$ where $L$ is the wire length. \cite{Bardarson07,Nomura07}  In contrast long non-topological wires invariably manifest a decreasing conductance.  These arguments based only on scattering and on spin-momentum locking give a detailed microscopic picture of how topology protects  conduction.

How does topological protection work when boundary disorder becomes large and both bulk and boundary states are included?  Recently Schubert et al \cite{Schubert12} performed a numerical simulation of 3-D TIs with boundary disorder and found that  the surface band evolves through three regimes when the boundary disorder is increased.  At small disorder the surface states are close to plane waves and their spectral density follows a linear Dirac dispersion.  However when the disorder grows larger  the in-gap states are no longer homogenous plane waves and therefore the linear Dirac relation between momentum and energy  disappears.   This is consistent with several experimental ARPES studies on 3-D TIs which saw the surface band signal weaken and blur as boundary disorder was increased. \cite{Hsieh09,Noh11,Liu12}    Lastly at very large boundary disorder the in-gap states move just inside the disordered region and return to the  Dirac dispersion found at weak disorder.   Based on these numerical results,  Schubert et al argued that  when the disorder is larger than the gap the bulk and surface states mix, the Dirac cone is destroyed,  and the TI surface states are no longer topologically protected.
 
The present paper studies the same physics, but in 2-D TIs, where the quantized $2 G_0$ edge conductance simplifies analysis, much larger sample sizes are accessible,  and statistical errors and finite size effects are more easily controlled.  We confirm that there are three disorder regimes, and that at very large disorder the in-gap edge states shift to the boundary between the clean bulk and the disordered edge. Like Schubert et al, we show that in the intermediate disorder regime the  edge states are inhomogenous (not plane waves) and therefore do not follow any dispersion relation between energy and momentum.  However we also show that edge conduction is topologically protected at any disorder strength - the $2 G_0$ quantization is never destroyed by boundary disorder.   In this intermediate disorder regime direct measurements of the spectral density will find no signal of the Dirac cone, while transport experiments will show robust edge conduction.

Going further, we find that the inhomogeneous in-gap states at intermediate disorder are controlled by a resonance between the disorder potential and the bulk bands.  Both the resonance and the accompanying edge state inhomogeneity  occur at the characteristic energy scale of the bulk bands; the in-gap states' plane-wave character is protected by the bulk band width, not by the much smaller bulk band gap.   The main effects of boundary disorder in the resonant region are to increase the in-gap density of states, decrease the Fermi velocity, and (if spin is conserved) pull the edge states into the disordered outer layers.  Since a TI's boundary easily can be  manipulated and measured, these effects may have useful applications to  TI device engineering.

 The structure of this paper is as follows.  In section \ref{Methods} we describe our model of a 2-D TI and the numerical methods which we use to study it, in section \ref{Results} we present and discuss our numerical results, and in section \ref{Discussion} we discuss the interpretation, and physics, and possible applications of these results.

 \section{Methods \label{Methods}}

   \begin{figure*}[]
\centering
\includegraphics[scale=1.3,clip]{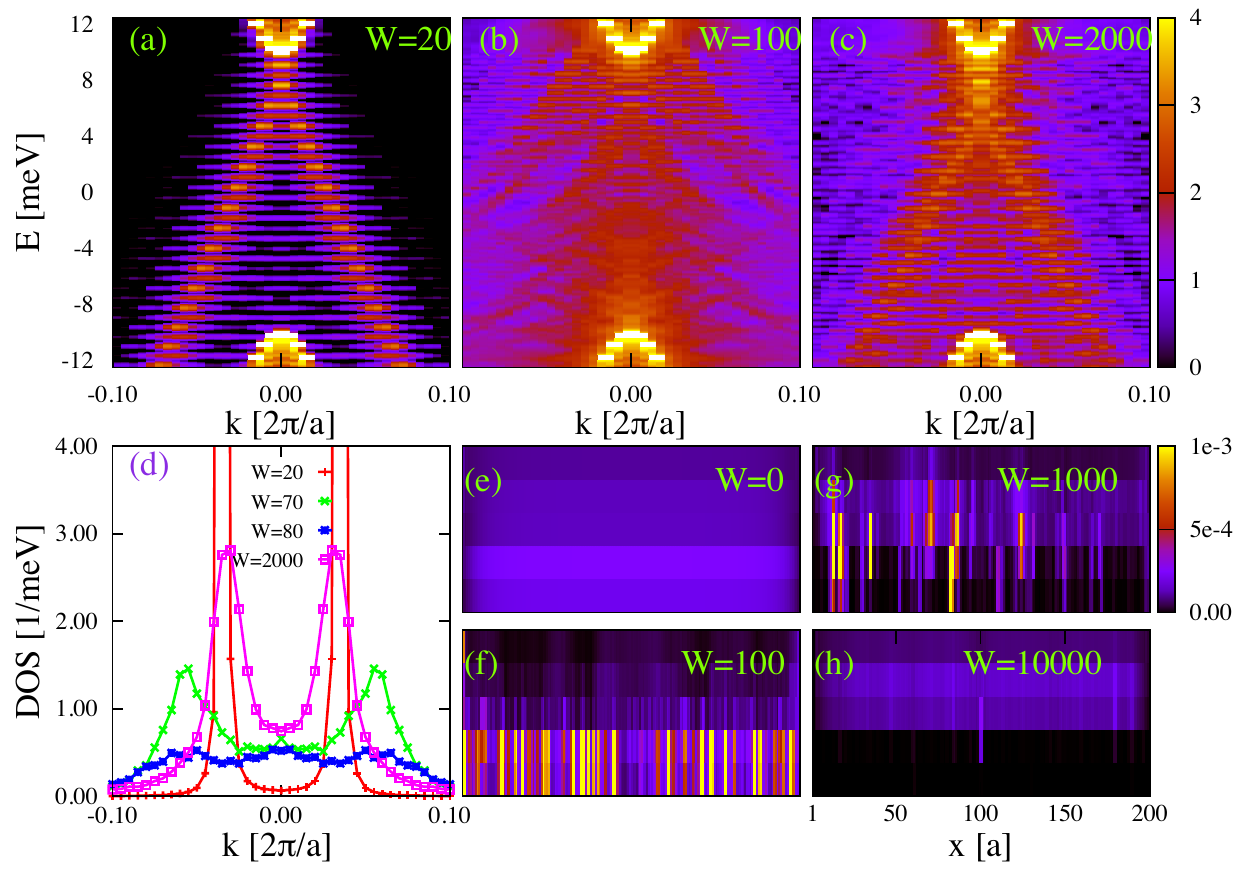}
\caption{ (Color online.) Effects of boundary disorder on the spectral density $A(\vec{k}, E)$ and typical in-gap edge states.  Panes a, b, and c show the spectral density as a function of both energy $E$ and wave-number $\vec{k} = (k, 0)$ at small disorder ($W = 20$ meV), intermediate disorder ($W = 100$ meV), and large disorder ($W = 2000$ meV.)    The  Dirac cone's two lines are clearly visible  at small disorder and large disorder, but disappear in a range starting at $W = 80$ meV and extending to near $W = 1100$ meV.  Pane d shows the averaged spectral density on a cross section $E = -2.5$ meV; the Dirac cone is signaled by two symmetric peaks and is absent in a range starting at $W = 80$ meV.  Panes e-h show typical edge states inside the gap at four representative disorder strengths.  In all cases they are extended and conducting, but at $W=100, 1000$ meV they lose their plane wave character and therefore are not visible in the spectral density. }
\label{Diraccone1a}
\end{figure*}

We study the Bernevig-Hughes-Zhang (BHZ) tight binding model   \cite{Bernevig06a} of 2-D topological insulators.  This model was obtained by starting with the six-band Kane model of the $\Gamma_6$ and $\Gamma_8$ bands and then removing the $L$1 sub-band which splits off.  The remaining basis has four orbitals: the first two are $|s, S_z = 1/2\rangle, \; |p, S_z = 1/2\rangle$, and the last two have $S_z$ reversed. \cite{Zhang12}  The BHZ model's momentum representation is:
\begin{eqnarray}
\mathcal{H} & = & \begin{bmatrix} h(\vec{k}) & 0 \\ 0 & h^*(-\vec{k}) \end{bmatrix}
\nonumber \\ h(\vec{k}) & = &   (A/a)  \sigma_x \sin (k_x a) +    (A/a) \sigma_y \sin (k_y a) +  M \sigma_z  
\nonumber \\
&+& 2 (\sigma_z B/a^2 + I \, D/a^2) (2 - \cos(k_x a) - \cos(k_y a))
\nonumber \\
D   &=  & 512\, meV \, nm^2, \; B= 686 \, meV \, nm^2, 
 \nonumber \\
  A & = & 364.5 \, meV \, nm, \; a = 5 \, nm, \; M = -10 \, meV
\end{eqnarray}
\noindent $\sigma_{x,y,z}$ are the Pauli matrices, and the sines and cosines in the Hamiltonian ensure the correct  topological band structure. $a = 5$ nm is the lattice spacing.  The penetration depth of the in-gap edge states is always less than $5 a = 25$ nm, which gives us good control of finite size effects when calculating the conductance in strips of width $3$a to $28$a.  Finite size effects are practically absent in our other calculations, where we have checked that changing the system size from $200$a $\times\; 200$a  to $400$a $\times \; 400$a  makes very little difference in our results.   The mass $M$ can be tuned by changing the sample thickness, and we choose a negative mass $M = -10$ meV so that there is a band inversion. This mass can produced by fabricating a quantum well with a thickness of  $\sim70$  \AA.   \cite{Rothe10}  With this mass the bulk exhibits a band gap in the interval $[-10,10]$ meV, and the Dirac point lies near $8$ meV.  The valence band extends over  $[-62, -10]$ meV, and the conduction band extends over $[10, 373]$ meV.  The bulk bandwidths of $52$ meV, $363$ meV are important - we will show that they set the energy scale where the edge states lose their plane-wave character.

The BHZ model  conserves the $S_z$ component of the spin, which is protected by  bulk inversion symmetry.  This symmetry can be broken by a quantum well or gate electrode \cite{Rothe10}, and also can be broken by disorder.  We introduce disorder on the outer two layers ($2$a = $10$nm) of the TI, and this disorder is white noise without any correlation between sites.  In the numerical results presented here we follow several papers \cite{Li09, Jiang09, Groth09}  which preserved the $S_z$ symmetry and  used nonmagnetic on-site disorder.  On each individual site the disorder is proportional to the identity and its strength is chosen randomly from the interval $\left[ -W/2, \, W/2 \right] $, where $W$ is the disorder strength.       With this $S_z$ - conserving disorder the BHZ model factorizes into independent $S_z = 1/2$ and $S_z = -1/2$ sectors which are time-reversed exact copies of each other.  Taken individually, each sector describes a Chern insulator with unit conductance $G_0$ in the bulk gap.  \cite{Haldane88}  Most of the  numerical results presented here, including all of our figures,  concern themselves only with the $S_z = 1/2$ sector of the BHZ model, and therefore apply equally to Chern insulators and to topological insulators.  We also have repeated many of our calculations using the full BHZ model and disorder which the breaks $S_z $ symmetry while preserving time reversal symmetry, and we will discuss these results briefly where they are relevant.   Although $S_z$ symmetry breaking is pertinent to disorder effects in the bulk  \cite{Groth09, Prodan11, Onoda07, Obuse07, Yamakage11} because the full BHZ model belongs to the symplectic universality class while each Chern insulator subsector belongs to the unitary class, this symmetry breaking is less important for  the present paper's results on boundary disorder with a clean bulk.  Here  the  conducting states  in the gap are edge states whose average behavior is independent of the BHZ model's $S_z$ symmetry.  

 With the exception of our conductance data, all of our numerical results come from a  leads-free geometry.  However for the conductance we use a strip with two attached clean semi-infinite leads that have widths equal to the width of the strip itself.  We evaluate the conductance using the Caroli formula\cite{Caroli71,Meir92} $G = -\frac{e^2}{h}{Tr}((\Sigma^r_{L}-\Sigma^a_{L}) G^r_{LR} (\Sigma^r_{R}-\Sigma^a_{R}) G^a_{RL})$, where $G^a, G^r$ are the advanced and retarded Green's functions connecting the left and right leads and $\Sigma_{L,R}$ are the self-energies of the leads. We evaluate the lead self-energies using the well-known iterative technique developed by Lopez Sancho et al. \cite{Lopez84}    
 
 \section{Results\label{Results}}

Panes a-d of Figure \ref{Diraccone1a} summarize the spectral density - the square $A(\vec{k}, E) = \sum_n \delta(E-E_n) \, |\langle \vec{k} | \psi_n \rangle |^2$ of the eigenstates $ |\psi_n \rangle$, resolved in momentum space. This quantity is very attractive because in pure samples it shows a clear X-shaped signal  marking the boundary state Dirac cone, and because in 3-D TIs it may be measured with ARPES.   Such measurements have been very popular both for proving that particular materials are topological insulators and for diagnostic evaluations of individual experimental samples.    In panes a-c of  Fig. \ref{Diraccone1a} we report the spectral density in the band gap of a single 2-D $400$a $\times \;400$a sample with periodic and open boundary conditions along the $x$ and $y$ axis respectively, using a logarithmic (base 10) color scale.   Pane a shows the spectral density of a weakly disordered sample ($W = 20$ meV), which is substantially the same as the density of a disorder-free sample. The two straight lines of the Dirac cone are very clear, and the  Dirac point lies near $8$ meV.   Above $10$ meV and below $-10$ meV the bulk conduction and valence bands are visible as very bright curves.    In pane b we increase the disorder strength to $W = 100$ meV and observe that the Dirac dispersion is no longer visible while the bulk bands remain unchanged.  Lastly pane c  shows that when the  disorder strength is far larger than the band widths ($W = 2000$ meV vs. $[52, 363]$ meV) the edge states  again manifest themselves in a clearly visible linear Dirac  dispersion.  Pane d tells the same story with more precision by focusing on a cross-section of $A(\vec{k} = (k,0), E)$ at $E = -2.5$ meV,  and averaging over $1000$ $200$a $\times\;200$a samples.  Cross-sections at other values of $k_y$ and $E$ are similar. The data for small disorder ($W = 20, 70$ meV) show two peaks at $|\vec{k}| = |E-E_0|/\hbar v_F$ corresponding to the two branches of the Dirac cone.  The peaks disappear at $W = 80$ meV  and reappear near $W = 1100$ meV.   These results agree completely with Schubert et al's study of the spectral density in 3-D TIs \cite{Schubert12} which found that the Dirac cone is absent from the spectral density at intermediate disorder but reappears at large disorder.  Our results also  are  consistent with several experimental ARPES observations of a weakened surface band signal in 3-D TIs with boundary disorder. \cite{Hsieh09,Noh11,Liu12}  

\begin{figure}[t]
\includegraphics[scale=0.5,clip]{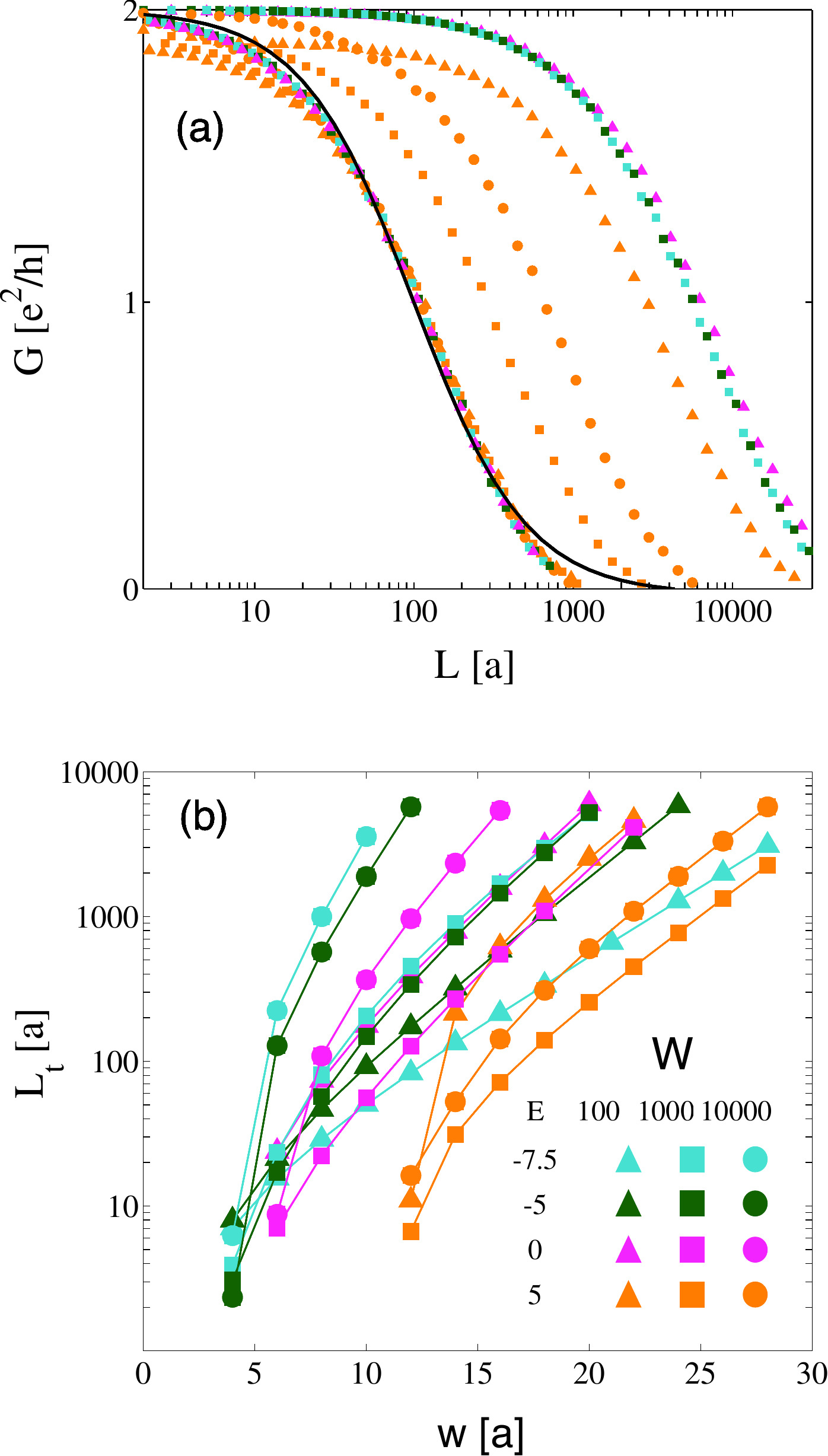}
\caption{(Color online.) Robust topological protection of  conduction, limited only by tunneling through the bulk.  Pane a shows how we determine the tunneling length $L_t$.  We plot the conductance as a function of wire length $L$ at six different disorder strengths and in-gap Fermi levels according to the legend in pane b.   In each case the average conductance  follows a universal curve, which is shown in the left-most curve.  The thin black line shows a best fit to the hyperbolic tangent function.  We determine the tunneling length $L_t$ by finding the point where the average conductance is equal to $1$.    Pane b shows $L_t$ as a function of strip width $w$ for several  in-gap Fermi energies and disorder strengths, including $W = 10^4$ meV which  far exceeds the band width.  Each curve converges rapidly to a straight line, proving that the tunneling length becomes exponentially large when the strip is widened, and that  conduction is topologically protected in wide strips.    }
\label{TunnelingSignature}
\end{figure}

 \begin{figure*}[t]
\centering
\includegraphics[scale = 1.4, clip]{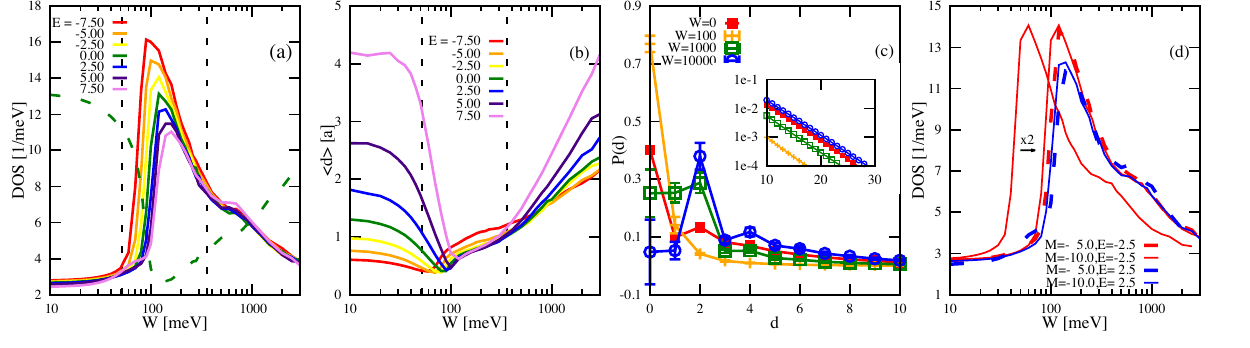}
\caption{(Color online.) The resonance between disorder and the bulk band which is responsible for disrupting the in-gap edge states' plane-wave character.  Pane a shows the density of states at seven Fermi levels inside the band gap.  A clear resonance is visible  at the characteristic bulk band widths $[52, 363]$ meV, which are marked with vertical dashed lines.  The green dashed line plots the profile of the Fermi velocity at $E = 0$ meV, which is inversely proportional to the DOS.  The units are arbitrary. Pane b shows the average distance from the edge $\langle d \rangle$, while pane c plots the probability $P(d)$ as a function of distance from the edge.  Both panes show that in the resonant region the edge states are pulled into the disordered layers.  The inset in pane c shows that deep inside the bulk the edge states are controlled by the penetration length plus a shift caused by the resonance.  Pane d shows the density of states when the band gap lies in the interval $[-10, 10]$ meV, or instead in the interval $[-5, 5]$ meV.  The curve far to the left shows where the $[-5, 5]$ meV DOS curve would be if the resonance were controlled by the band gap instead of the band width. }
\label{DOSRenormalized}
\end{figure*}

To understand the fate of the in-gap edge states we calculate individual edge states at four representative disorder strengths and show their probability distributions $|\psi(\vec{x})|^2$ in panes $e-h$  of Figure \ref{Diraccone1a}.  We plot the lower  $5$ layers of $200$a $\times\, 200$a samples with open boundary conditions, choose $E = 7.5$ meV, and employ a linear color scale.    In all cases the great majority of edge states remain extended along the entire perimeter of the TI sample and strongly avoid the sample's interior.   Pane e shows an edge state at zero disorder.  It is a plane wave; its density  is constant along the sample edge.   Likewise pane h shows that at  large disorder ($W = 10^4$ meV) the edge states shift inward into the sample, avoiding the outer two disordered layers, and therefore have a  smooth plane wave character.  However at intermediate disorders  $W = 100, 1000$ meV the  edge states are strongly affected by the disorder and become very  inhomogeneous.  This  explains the absence of a Dirac cone structure in the spectral density.  Extended edge states are present inside the gap regardless of the disorder strength, but at $W = 100, 1000$ meV they are not plane waves and therefore have no structure  in momentum space.   

We have checked thoroughly that  the majority of  in-gap states are spatially extended 1-D states propagating near the TI edge.  This is proved by the eigenvalues  which remain very uniformly spaced at all disorder strengths and all in-gap Fermi  levels, as is true of 1-D delocalized states.    Moreover we studied the size distribution of 30,000 in-gap edge states in  $200$a $\times \,200$a samples with open boundary conditions at $E = 5$ meV, and our statistics exclude the existence of any spatially extended bulk states inside the gap.   For comparison, in a clean $W=0$ sample the two states bracketing $E = 5$ meV have PRs equal to $4850, \, 5130$, and these states are certainly edge states because $E = 5$ meV lies inside the bulk.  Extended bulk states would have PRs of the order of the system volume $V = 80,000$.  In the presence of boundary disorder the in-gap PR is always reduced compared to the clean PR of $5130$, proving that the in-gap states are not bulk states and confirming that disorder reduces their penetration into the sample.  At $W = 100$ meV $1\%$ of the states have $PR \geq 395$, $0.1\%$ have $PR \geq 422$, and the  average and maximum PRs are $276, \, 465$.   At $W = 1000$ meV $1\%$ of the states have $PR \geq 713$, $0.1\%$ have $PR \geq 891$, and the average and  maximum PRs are $261, \,1070$.  At $W = 10,000$ meV $1\%$ of the states have $PR \geq 4246$, $0.02\%$ have $PR \geq 5000$, and the average and maximum PRs are $1480, 5114$.   Our statistics do reveal a small sub-population of very localized states in the gap, but  these states are a small fraction of the total.    At $W = 10^3$ meV $11.9 \pm 0.2$ percent of the  in-gap states have participation ratios (PRs) less than $50$, and $2.4 \pm 0.1$ percent have PRs less than $10$.  At $W = 10^4$ meV these proportions are $5.9 \pm 0.1 $ and $4.9 \pm 0.1$ percent, and at $W = 10^2$ meV the proportions are much smaller.   In summary, the in-gap states do not reside in the bulk and almost all are extended around the sample perimeter.

%  $E = 5$    E = 7.5 + 0.171: PR = 0.000128. E = 5, 20 eigienvalues: E = 5 -3.593: 0.000315; E = 5 -0.526: 0.000218; E = 5 - 0.136: 0.000206; 4850 E = 5 + 0.255: 0.000195; 5130 E = 5 + 3.945: 0.000100.  The PRs change monotonically and smoothly over this interval of 20 eigenvalues.  

%\footnote{In $200$a $\times \,200$a samples at $E = 5$ meV $21.2 \pm 0.4$ percent of the states have participation ratios (PRs) less than $100$, and $11.9 \pm 0.3$ percent have PRs less than $50$ 2.4 \pm 0.2 below 10.  At $W = 10^4$ meV these proportions are $6.7 \pm 0.8 $ percent and $5.9 \pm 0.8 $ percent.  4.9  \pm 0.7 below 10 } 

Although the edge states are no longer visible in the spectral density, they remain topologically protected and  in wide strips the in-gap conductance remains forever quantized.  In Figure \ref{TunnelingSignature} we calculate the conductance in  long TI strips with open boundary conditions and show that the only mechanism which destroys conduction is tunneling across the strips. In these conductance calculations we introduce disorder only in the outermost layer of the TI. The tunneling occurs at a characteristic strip length $L_t$ which can be measured by calculating the average conductance as a function of strip length $L$, as seen in Fig. \ref{TunnelingSignature}a.   We calculated this function at three disorder strengths $W = \{100, 1000, 10000\}$ meV and four  in-gap Fermi energies $E_F = \{ -7.5, -5, 0, 5\}$ meV,  averaging over 3000 disorder realizations  and keeping the strip width fixed at $w  = 20$a.  In each case (except for small strip widths) the  average conductance  follows an identical curve, as shown in the left-most curve where we have shifted all the data to coincide with each other \footnote{Small deviations from universality are seen in the orange $E_F = 5$ meV data, which never reaches $G = 2 G_0$.  This is a finite size effect.   The conductance  is suppressed at $E_F = 5$ meV in thin strips because a gap opens around  the Dirac energy $E_F \approx 8$ meV. }.  The universal curve is nearly linear in the interval $G = [0.4, 1.4]$ and is fairly similar to the hyperbolic tangent, which is marked by a thin black line.  This makes us confident that we can reliably calculate the tunneling length scale $L_t$ by determining the value of $L$ which satisfies $G(L) = 1$.  In pane b we plot $L_t$ as a function of strip width $w$.  Each curve is determined by varying the strip width while keeping  the disorder strength and Fermi level constant.  We explore the bulk band gap between $[-10,10]$ meV by calculating four Fermi energies $E = -7.5, -5, 0, 5$ meV.  We also explore three disorder strengths $W = 10^2, 10^3, 10^4$ which all far exceed the bulk band gap, and the last of these far exceeds the band width.   At small strip widths the curves show some bending caused by finite size effects.  Nonetheless  each curve quickly converges to a straight line;  the tunneling length $L_t$ is proportional to $L_t \propto e^{w/w_t}$ for some constant  $w_t $, and is exponentially large in a wide strip \footnote{The straight lines relating the log of the tunnelling length $\ln L_t$ to strip width  $w$ generally have slopes around $0.10 - 0.15$, but we find slopes about double this value at $W = 10^4, \; E_F = -7.5, \, -5.0$. The y-intercepts ($w = 0$ values) of these straight lines all lie between $~1.5$ and $~20$.}.    This proves that the conductance remains forever quantized in an infinitely wide strip;  conduction is robustly protected regardless of the disorder strength.  We find similar but much weaker tunnelling behaviour when using disorder that breaks the BHZ model's $S_z$ symmetry.  Both with and without $S_z$ symmetry the edge states inside the bulk band gap remain always extended and conducting.

Lastly we study the physics which disrupts the Dirac cone signal in the spectral density, which is a resonance between the disorder potential and the bulk bands at the characteristic bulk band widths $[52, 363]$ meV.   In Figure \ref{DOSRenormalized} we study $200$a $\times\; 200$a squares with open boundary conditions.  Pane a reports the density of states (DOS)'s dependence on disorder, which we obtained by  collecting the three eigenvalue-eigenstate pairs nearest to specified Fermi energies $E = -7.5, ..., 7.5$ meV and averaging over $100$ disorder realizations.  First we note that the DOS (and all other observables) evolves continuously; the absent Dirac cone signal  can not be explained by an absence of in-gap edge states.  Inside the interval $W = [52, 363]$ meV the DOS first grows sharply and then begins to decay more slowly.  The peak over base ratio of this resonance   is about $5.8$.  In the left shoulder of the resonance we see a spread between the DOS curves measured at different Fermi energies, indicating that the density of states increases first near the valence band and soon after at higher Fermi energies.  This suggests that at around $W \approx 52$ meV the valence band begins to shed states into the gap, and that as the disorder is increased these states move further into the gap and later are joined by states from  the conduction band.   The valence band has a much smaller bandwidth, so it is natural that relatively weak disorder  $W \approx 52$ meV would affect only the DOS near the valence band, while considerably stronger disorder affects the whole spectrum. Outside of the resonance the DOS tends toward its disorder-free value. Except for in the interval $W \approx [52, 200]$ meV,    the DOS is  roughly  independent of the Fermi energy, as expected of Dirac states on the edge of a 2-D TI.   
 
The above discussion suggests that  the bulk band width not the bulk band gap controls the energy scale at which the in-gap edge states lose their plane-wave character.  This result contrasts with at least one previous paper \cite{Schubert12} which expected that the Dirac cone is protected by the bulk band gap.  Therefore we checked this question directly by changing the gap to $[-5, 5]$ meV, in contrast to the rest of our calculations where we set the gap to $[-10, 10]$ meV.  We recomputed the DOS, the average distance from the edge, and the participation ratio.   If the bulk gap were the controlling factor then a $0.5$ reduction of the band gap should shift all observables to smaller disorder, i.e. $W \rightarrow W/2$.   Pane d of Figure \ref{DOSRenormalized} shows where the DOS would shift to if the bulk gap were the controlling scale.  We observe that the DOS (and other observables) shift slightly in the opposite direction.   This confirms that the boundary states' plane-wave character is protected against boundary disorder by the bulk band width not the bulk band gap, and agrees with the well-established fact that in 2-D TIs bulk disorder destroys the topologically protected edge band only when it reaches the scale of the bulk band. \cite{Groth09,Xu12}

 Panes b and c of Figure \ref{DOSRenormalized} show that the resonance pulls the edge states into the disordered sites on the edge of the TI. Two authors have already briefly discussed this behavior at a qualitative level \cite{Chu09,Jiang09}; here we measure it quantitatively and establish its dependence on the disorder strength and the Fermi  energy. Pane b plots the average distance from the edge - the expectation value $\langle d \rangle = \int d^2{\vec{x}}s |\psi(\vec{x})|^2 d(\vec{x})$  of the distance $d(\vec{x})$ to the nearest edge.   Again we have averaged over $100$ disorder realizations.   Outside of the  resonant interval $[52, 363]$ meV  $\langle d \rangle $  depends strongly on the Fermi level and is maximized near the Dirac point $E = 8$ meV.  This behavior is linked to the penetration depth which is also energy dependent, and contrasts strongly with the resonant interval where $\langle d \rangle$ is  practically independent of energy.  Inside the resonance  the state's position is determined by disorder and not  by the penetration depth.   $\langle d \rangle $ is strongly reduced to  about one lattice unit; the edge states reside   almost entirely in the outer two layers where we have introduced disorder.  We see this more precisely in pane c, which plots the probability $P(d)= \int d^2{\vec{x}} |\psi(\vec{x})|^2 \, \delta(d - d(\vec{x}) )$ that the state is located on layer $d$. We examine four disorder strengths, average over $10^4$ disorder realizations, and keep the Fermi energy fixed at $5$ meV.   The error bars plot the state-to-state standard deviation of $P(d)$; this statistic's smallness indicates that  most states share the same $P(d)$.  Inside the resonant interval $P(d)$ becomes strongly concentrated on the outer two layers; the probability on the outer layer $P(0)$ almost doubles from the pure case  to $W = 100$ meV.  As the disorder is increased beyond the band width the in-gap edge states are expelled from the disordered layers which are no longer part of the TI because their band structure is overwhelmed by disorder.  The edge states  concentrate at the boundary between the disordered layers and the clean  layers which retain their TI band structure.  The inset shows that deep inside the TI  $P(d)$ is always controlled by the penetration depth and decays exponentially. Here the only effect of disorder is to shift $P(d)$ first inward  at intermediate disorder strengths and then outward at large disorder.  This effect is similar to the  phase shifts found in the theory of scattering events.
 
We do not observe the in-gap edge states being pulled into the disordered region when we use disorder  that breaks the BHZ model's $S_z$ symmetry.  In this case the average distance from the edge  $\langle d \rangle $ does not decrease inside the resonance, and always retains its dependence on the Fermi level.  The resonance in the DOS is  a very symmetric and smooth peak centered  around $E \approx 300$ meV, and its peak over base ratio is half the value found with $S_z$ conserving disorder.  Although we expect both conductance quantization and the resonance to be robust against changes in the Hamiltonian and in the disorder type as long as time reversal symmetry is preserved, the particulars of the resonance and its effects on the in-gap edge states are more sensitive to these details.

\section{Discussion \label{Discussion}} 

The resonance which we have observed in the DOS is caused by resonances between   individual disordered sites and the in-gap edge states.  As an electron moves along the edge of the TI it from time to time becomes almost trapped at a particular site and dwells there for a while before continuing its journey.  This trapping is unable to destroy the in-gap edge state but it does introduce both a phase shift and a localized increase in the edge state's probability density.  This destroys both the edge state's plane wave character and the Dirac cone signal in the spectral density. The trapping naturally also decreases the Fermi velocity  $v_F$ of the edge states.  Both dimensional analysis and the Dirac relation $E = v_F \hbar |k|$ indicate that $v_F$ is inversely proportional to the density of states, so in  pane a of Figure \ref{DOSRenormalized} we have plotted the inverse of the DOS, which reveals the profile of the Fermi velocity.  At its peak the resonance divides the (average) Fermi velocity by $5.8$ and as a result multiplies the DOS by the same number.

In synthesis, our results  suggest that even though there is no dispersion relation for individual realizations of disorder, \textit{on average} the  in-gap edge states always remain controlled by a one-dimensional two-band effective edge  Hamiltonian, albeit renormalized by disorder.    To be clear, here we are discussing disorder-averaged properties of the in-gap states, and  an effective Hamiltonian describing those averages - not  the real properties of individual eigenstates and eigenvalues.   In particular, the robust  $2 G_0$ conductance indicates that the edge states never backscatter, which is consistent with spin-momentum locking, a general feature of $T$ preserving two-band Hamiltonians with strong spin-orbit couplings.
Moreover we have found that the edge states' energy levels always retain the level repulsion and rigid spacing that signal 1-D extended states.        Lastly with the exception of the interval $W \approx [52, 200]$ meV the DOS is roughly independent of the Fermi energy as expected of a linear 1-D Dirac Hamiltonian;  the effective Hamiltonian preserves the Dirac Hamiltonian's structure with renormalized parameters.  The main effects of boundary disorder on the TI's in-gap disorder-averaged properties are to reduce the Fermi velocity, increase the DOS, pull the edge states into the disordered layers if spin is conserved, and at large disorder shift the edge states to the boundary with the clean bulk.  

Our observations  of a reduced Fermi velocity and increased DOS  in the resonance duplicate the behavior of  non-disordered TIs with soft boundary conditions; i.e. models where the TI boundary is created by a slowly varying potential.   \cite{Stanescu09,Buchhold12}  In fact it is well known that for calculations of the DOS the main effects of a disordered potential can be duplicated using a spatially uniform effective potential; \cite{Soven67} our introduction of disordered surface layers can be understood as introducing an effective potential which widens the transition region at the TI boundary.  It is therefore natural to speculate that the in-gap density of states at the resonance peak might be made quite large by increasing the number of disordered layers.  If this effect were duplicated in 3-D TIs it may be of great utility for increasing the surface conductivity.   Moreover TI devices with a patterned Fermi velocity and refractive index may be fabricated by applying disorder to selectively to specific regions on the TI surface, and such devices may display interesting lensing and reflection effects. \cite{Moon11}  Further investigation of these speculations and applications are left to future works.

Quansheng Wu and Liang Du contributed equally to this work.  We gratefully acknowledge discussions with and support from Xi Dai, Zhong Fang, Sheng Nan Zhang, Tomi Ohtsuki, Koji Kobayashi, Dimi Culcer, Markus Mueller, and Ivar Martin.  We thank the reviewers for many helpful suggestions, and especially for the connections to Chern insulators and to TIs with soft boundary conditions. This work was supported by the National Science Foundation of China and by the 973 program of China  under Contract No. 2011CBA00108.

%\begin{appendix}
\bibliography{Vincent}
\end{document}